\documentclass[twocolumn,aps,pra,showpacs,superscriptaddress,floatfix]{revtex4}

\usepackage{bm}
\usepackage{dcolumn}
\usepackage{graphicx}
\usepackage{epsf}
\usepackage{epsfig}
\usepackage{nicefrac}
\usepackage{times}

\usepackage{amsmath}
\usepackage{amsfonts}
\usepackage{amssymb}

\newcommand{\addrFR}{Theoretische Quantendynamik, 
Physikalisches Institut der Universit\"{a}t Freiburg,\\
Hermann--Herder--Stra\ss{}e 3, 79104 Freiburg im Breisgau, Germany}

\newcommand{\addrNIST}{National Institute of Standards and Technology,
Gaithersburg, Maryland 20899--8401}

\begin{document}

\bibliographystyle{myprsty}

\title{Self--Energy Correction to the Two--Photon Decay Width in
Hydrogenlike Atoms}

\author{Ulrich D.~Jentschura}
\affiliation{\addrFR}
\affiliation{\addrNIST}

\begin{abstract}
We investigate the gauge invariance of the leading 
logarithmic radiative correction to the two-photon decay width 
in hydrogenlike atoms. It is shown that an effective treatment
of the correction using a Lamb-shift ``potential'' leads to 
equivalent results in both the length as well as the velocity gauges
provided all relevant correction terms are taken into account.
Specifically, the relevant radiative corrections are related to the 
energies that enter into the propagator denominators, to the Hamiltonian,
to the wave functions,
and to the energy conservation condition that 
holds between the two photons; the form of all of these effects
is different in the two gauges, but the final result is shown to 
be gauge invariant,
as it should be. Although the actual calculation
only involves integrations over nonrelativistic hydrogenic Green functions,
the derivation of the leading logarithmic correction can be regarded
as slightly more complex than that of other typical logarithmic terms.
The dominant radiative correction to the $2S$
two-photon decay width is found to be
$-2.020\,536\, (\nicefrac{\alpha}{\pi})\,(Z\alpha)^2\,\ln[(Z\alpha)^{-2}]$
in units of the leading nonrelativistic expression. This result is
in agreement with a length-gauge calculation [S.~G.~Karshenboim and
V.~G.~Ivanov, e-print physics/9702027], where the coefficient
was given as $-2.025(1)$.
\end{abstract}

\pacs{12.20.Ds, 31.30.Jv, 06.20.Jr, 31.15.-p}

\maketitle

%
%
\section{INTRODUCTION}
\label{intro}

The two-photon decay of the metastable $2S$ level in atomic
hydrogen and hydrogenlike systems is a rather intriguing physical phenomenon;
it was first investigated by
M. G\"{o}ppert--Mayer a long time ago~\cite{GO1929,GM1931}.
The transition involving two quanta limits the lifetime of the 
metastable $2S$ resonance, at least for low and medium
nuclear charge numbers $Z$. By contrast, the highly suppressed 
magnetic dipole transition to the $1S$ ground state
has a negligible influence on the decay width~\cite{MaMo1978}.
In this article, we evaluate the dominant 
self-energy radiative correction to the two-photon process.
We recall here the 
known leading-order result~\cite{ShBr1959,ZoRa1968,Kl1969,remark2S}
\begin{equation}
\label{etauold}
\tau^{-1} \approx \Gamma_0 = 8.229 \, Z^6\, {\rm s}^{-1} = 
1.310 \, Z^6\, {\rm Hz}\,.
\end{equation}
For ionized helium ($Z=2$),
rather accurate experimental verifications
of this result exist~\cite{Pr1972,KoClNo1972,HiClNo1978}.
Due to its metastability, the 2S level in hydrogenlike systems
is one of the most accurately defined 
resonances found in nature. Indeed, it is this very property---the small
natural linewidth---which has made possible the high-resolution two-photon 
spectroscopy of the $1S$--$2S$ 
transition~\cite{UdEtAl1997,HuGrWeHa1999,ReEtAl2000,%
NiEtAl2000,BeEtAl1997,BeEtAl2000}.

The fully relativistic quantum electrodynamic formalism is 
intricate when applied to bound-state 
problems~\cite{Sc1961,ItZu1980,CaLe1986,Ho2004}, 
but it is often possible to gain 
a rather good understanding of QED radiative corrections
to a particular process if one uses a simplified, NRQED Lagrangian
that contains effective operators 
which then lead to the perturbations that have to be 
evaluated (see e.g.~\cite{KiNi1996,NiKi1997}).
Of course, the main difficulty of any bound-state calculation,
which is
the separation of the two energy scales (scale of binding energy
and the energy/mass scale of the free particles), persists in 
the effective approach. It is necessary to also specify 
cutoff prescriptions; the artificially introduced scale-separation 
parameters then cancel at the end of the 
calculation~\cite{Pa1993,JePa1996,KiNi1996,NiKi1997}.
Elucidating discussion of the latter point 
can be found 
in~\cite[Ch.~123]{BeLiPi1989} and in~\cite[Ch.~11.4 on p.~493]{We1995i}.

Within nonrelativistic 
quantum electrodynamics (also referred to 
as NRQED, see~\cite{CaLe1986,Ho2004}), one has the choice 
between two different forms of the interaction Hamiltonian: the 
``length'' (Yennie) and the ``velocity'' 
(Coulomb) gauges. There are certain intriguing 
issues involved with the gauge invariance in the dynamical nonrelativistic
atom-light interaction. Indeed, in order to prove gauge invariance for 
dynamical processes, it is in many cases necessary
to carefully consider the gauge transformation
of the atomic wave function in addition to the transformation of the fields.
Otherwise, non-gauge invariant results are obtained off 
resonance~\cite{Ko1978prl,ScBeBeSc1984,LaScSc1987}. 
In the current situation of radiative corrections to the 
two-photon decay width, we will show that it is possible to 
ignore the transformation of the wave function: the two-photon decay width,
including the radiative corrections, is invariant 
under a ``hybrid'' gauge transformation~\cite{ScBeBeSc1984}
which involves only the fields, but ignores the gauge transformation
of the wave function. In general, the choice of the 
gauge and the interpretation of physical operators 
have to be considered very carefully 
in time-dependent problems (see~\cite[p.~268]{La1952}
and Refs.~\cite{Ko1978prl,ScBeBeSc1984,LaScSc1987}). 

The gauge invariance of the
two-photon decay rate and of the radiative corrections
to this effect 
can be regarded as slightly problematic, partly because the
integration over the photon energy is restricted to a finite
interval. By contrast, the gauge invariance of the low-energy
part of the one-loop
self-energy shift, in an effective NRQED treatment, holds only
because one may drop terms whose divergence, for large
photon frequency, is stronger
than logarithmic~\cite[see Eq.~(3.4)~ff.]{Pa1993}; in this case
gauge invariance would be violated over finite intervals of the
virtual photon frequency.
It has been one of the main motivations for the current
paper to study related questions.

This article is organized as follows:
In Sec.~\ref{leading}, the leading nonrelativistic
contribution to the two-photon decay rate is discussed,
together with its relation to the NRQED two-photon
self-energy.
In Sec.~\ref{radcor}, the leading logarithmic radiative
correction to the two-photon decay rate is formulated,
the discussion is
based on a perturbation with an effective potential.
Explicit expressions are derived
in Secs.~\ref{length} and~\ref{velocity}
for the length and velocity gauges, respectively.
Gauge invariance is proven in Sec.~\ref{proof}.
Numerical results are presented in Sec.~\ref{numerical}.
Conclusions are drawn in Sec.~\ref{conclusions}.
All derivations are presented in some detail, for the
sake of transparency.

%
%
\section{LEADING--ORDER TWO--PHOTON DECAY RATE}
\label{leading}

The decay width of a bound system 
may be understood naturally as 
the imaginary part of the self energy~\cite{BaSu1978}.
Indeed, the (negative) imaginary part of 
the self-energy is just $\Gamma/2$,
where $\Gamma$ is the decay width. 
We discuss the derivation of the two-photon 
width based on this concept, within nonrelativistic quantum
electrodynamics~\cite{remarkNRQED}. 

The formulation of the two-loop self-energy problem 
within the context of nonrelativistic quantum 
electrodynamics (NRQED) has been discussed in~\cite{Pa2001}.
We denote by $p^j$ the Cartesian components of the momentum 
operator $\bm{p} = -{\rm i}\,\bm{\nabla}$.
The expression for the two-loop self-energy shift 
reads~\cite{Pa2001,remarkPRA2002}
\begin{scriptsize} 
\begin{widetext} 
\begin{eqnarray}
\label{NRQED}
\lefteqn{
\Delta E_{\rm NRQED} = 
- \left( \frac{2 \, \alpha}{3 \,\pi\,m^2} \right)^2 \,
\int_0^{\epsilon_1} {d}\omega_1 \, \omega_1 \,
\int_0^{\epsilon_2} {d}\omega_2 \, \omega_2 \, 
\left\{  
\left< p^i \, \frac{1}{H - E + \omega_1} \, p^j \, 
\frac{1}{H - E + \omega_1 + \omega_2} \, p^i \, 
\frac{1}{H - E + \omega_2} \, p^j \right> \right.} \nonumber\\[1ex]
& & + \frac{1}{2} \,
\left< p^i \, \frac{1}{H - E + \omega_1} \, p^j \,
\frac{1}{H - E + \omega_1 + \omega_2} \, p^j \, 
\frac{1}{H - E + \omega_1} \, p^i \right> \nonumber\\[1ex]
& & + \frac{1}{2} \,
\left< p^i \, \frac{1}{H - E + \omega_2} \, p^j \,
\frac{1}{H - E + \omega_1 + \omega_2} \, p^j \, 
\frac{1}{H - E + \omega_2} \, p^i \right>
\nonumber\\[1ex]  
& & + 
\left< p^i \, \frac{1}{H - E + \omega_1} \, p^i \, 
\left( \frac{1}{H - E} \right)' \, p^j \, 
\frac{1}{H - E + \omega_2} \, p^i \right> 
\nonumber\\[1ex]  
& & - \frac{1}{2} \,
\left< p^i \, \frac{1}{H - E + \omega_1} \, p^i \right> \,
\left< p^j \, \left( \frac{1}{H - E + \omega_2} \right)^2 \, p^i \right>
- \frac{1}{2} \,
\left< p^i \, \frac{1}{H - E + \omega_2} \, p^i \right> \,
\left< p^j \, \left( \frac{1}{H - E + \omega_1} \right)^2 \, p^i \right>
\nonumber\\[1ex]
& & \left. - m \,
\left< p^i \, \frac{1}{H - E + \omega_1} \, 
\frac{1}{H - E + \omega_2} \, p^i \right>
- \frac{m}{\omega_1 + \omega_2} \,
\left< p^i \, \frac{1}{H - E + \omega_2} \, p^i \right>
- \frac{m}{\omega_1 + \omega_2} \,
\left< p^i \, \frac{1}{H - E + \omega_1} \, p^i \right>
\right\}.
\end{eqnarray}
\end{widetext} 
\end{scriptsize} 
All of the matrix elements are evaluated on the reference state
$|\phi\rangle$, for which the nonrelativistic Schr\"{o}dinger wave
function is employed.
The expression for the two-photon
decay width [Eq.~(\ref{tpd1}) below] now follows in a natural
way as the imaginary part generated by the sum of the 
first three terms in curly brackets in
Eq.~(\ref{NRQED}). Specifically, the poles are generated
upon $\omega_2$-integration by
the propagator 
\begin{equation}
\frac{1}{H - E + \omega_1 + \omega_2} = 
\sum_{\phi'} \frac{| \phi' \rangle \, \langle \phi' |}
  {E' - E  + \omega_1 + \omega_2}
\end{equation} 
at $\omega_2 = E - E' - \omega_1$.
Alternatively, this condition may be expressed as
$E - E' = \omega_1 + \omega_2$,
and represents the energy conservation 
condition for the two-photon decay.
The imaginary part generated by the 
first three terms in curly brackets of the 
energy shift~(\ref{NRQED}) is thus seen to yield the two-photon
decay width~\cite{remarkONE}.

In view of the above discussion, and in agreement with 
Shapiro and Breit~\cite[Eq.~(3)]{ShBr1959}, 
the nonrelativistic expression for the 
two-photon decay width $\Gamma_0$ in the case $|\phi\rangle = | 2S\rangle$ 
and $|\phi'\rangle = | 1S\rangle$ reads
\begin{eqnarray}
\label{tpd1}
\Gamma_0 &=& \frac{4}{27}\,\frac{\alpha^2}{\pi}\,
\int\limits^{\omega_{\rm max}}_0 
d\omega_1 \, \omega^3_1\,\omega^3_2 \,
\nonumber\\[1ex]
& & \left| \left< \phi' \left| 
x^i \, \frac{1}{H - E + \omega_2} \, x^i \right| \phi \right> \right.
\nonumber\\[1ex]
& & \left.
+ \left< \phi' \left| 
x^i \, \frac{1}{H - E + \omega_1} \,
x^i \right| \phi \right> 
\right|^2\,,
\end{eqnarray}
where $\omega_2 = \omega_{\rm max} - \omega_1$
and $\omega_{\rm max} = E - E'$ is the maximum energy 
that any of the two photons may have.
When comparing this expression
to Eq.~(2) of~\cite{KaIv1997c}, it should be 
noted that the quantity $y$ {\em ibid.}
represents a scaled photon energy.
The Einstein summation convention is used throughout this article.
Note the following identity~\cite{BaFoQu1977,Ko1978prl}
\begin{eqnarray}
\label{tpd2}
\lefteqn{\left< \phi' \left|
\frac{p^i}{m} \, \frac{1}{H - E + \omega_1} \,
\frac{p^i}{m} \right| \phi \right> }
\nonumber\\[1ex]
& & 
+ \left< \phi' \left| 
\frac{p^i}{m} \, \frac{1}{H - E + \omega_2} \,
\frac{p^i}{m} \right| \phi \right> 
\nonumber\\[1ex]
&=&
-\omega_1\,\omega_2\,m^2\,
\left\{ \left< \phi' \left| 
x^i \, \frac{1}{H - E + \omega_1} \,
x^i \right| \phi \right> \right.
\nonumber\\[1ex]
& & \left.
+ \left< \phi' \left| 
x^i \, \frac{1}{H - E + \omega_2} \,
x^i \right| \phi \right> \right\}\,,
\end{eqnarray}
which is valid at exact resonance $\omega_1 + \omega_2 = E-E'$.
This identity permits a reformulation of the problem in the 
velocity-gauge as opposed to the length-gauge form.

%
%
\section{RADIATIVE CORRECTIONS}
\label{radcor}

We consider a hydrogenlike atom and employ natural units
with $\hbar = \epsilon_0 = c = 1$.
In order to analyze the 
radiative correction to the 
two-photon decay width,
one could write down all Feynman diagrams which contribute
to the process, and start evaluating them.
However, a much more economical understanding into the problem
can be gained by considering an approach inspired by
effective field theory, or nonrelativistic
quantum electrodynamics~\cite{CaLe1986,Ho2004},
in which the leading effect due to radiative photons is described by
an effective Lamb-shift potential~\cite{Ka1996,JeNa2002}
\begin{equation}
\label{radpot}
\delta V_{\rm Lamb} = \frac43\,\alpha\,(Z\alpha)\, 
\ln[(Z\alpha)^{-2}] \, 
\frac{\delta^{(3)}(\bm{r})}{m^2}\,.
\end{equation} 
In this work we will consider a ``standard normalized 
perturbative local potential''~\cite{Je2003jpa}
\begin{equation}
\label{standard}
\delta V = \frac{\pi (Z\alpha)}{m^2} \, \delta^{(3)}(\bm{r})\,.
\end{equation}
which is related to $\delta V_{\rm Lamb}$ by a simple prefactor,
\begin{equation}
\label{VLamb}
\delta V_{\rm Lamb} =
\frac43\,\frac{\alpha}{\pi} \, \ln[(Z\alpha)^{-2}] \, \delta V\,.
\end{equation}

The corrections to the Hamiltonian, to
the energy and to the wavefunction,
incurred by the perturbative potential~(\ref{standard}), read
as follows,
\begin{subequations}
\label{corrections}
\begin{eqnarray}
E &\to& E + \delta E \,, 
\\[1ex]
\delta E &=& \langle \phi | \delta V | \phi \rangle\,,
\\[1ex]
H &\to& H + \delta V  \,,
\\[1ex]
| \phi \rangle &\to& | \phi \rangle +
| \delta \phi \rangle \,,
\\[1ex]
| \delta \phi \rangle &=&
\left( \frac{1}{E - H} \right)' \,
\delta V | \phi \rangle\,.
\end{eqnarray}
\end{subequations}
The standard potential (\ref{standard})
leads to a ``normalized'' energy shift with unit prefactors,
\begin{equation}
\delta E (nS) = \frac{(Z\alpha)^4\,m}{n^3}\,.
\end{equation}
%

%
%
\section{LENGTH GAUGE}
\label{length}

According to (\ref{tpd1}), the two-photon 
decay rate $\Gamma_0$ of the metastable $2S$ state
is given by
\begin{equation}
\label{decaylength}
\frac{\Gamma_0}{A} = \int\limits_0^{\omega_{\rm max}}
d\omega_1 \, \omega^3_1 \, \omega^3_2 \, \zeta^2\,,
\end{equation}
where we use the definition
\begin{equation}
\label{defA}
A = \frac{4}{27}\,\frac{\alpha^2}{\pi}\,,
\end{equation}
as well as
$\omega_2 \equiv E_{2S} - E_{1S} - \omega_1$ and
$\omega_{\rm max} \equiv E_{2S} - E_{1S}$.
The quantity $\zeta$ is given by
\begin{subequations}
\label{defzeta}
\begin{equation}
\zeta = \zeta_1 + \zeta_2\,,
\end{equation}
where
\begin{eqnarray}
\label{defzeta1}
\zeta_1 &=& \left< 1S \left| x^i \, 
\frac{1}{H - E_{2S} + \omega_1} \,
x^i \right| 2S \right> \,,
\\[1ex]
\label{defzeta2}
\zeta_2 &=& \left< 1S \left| x^i \, 
\frac{1}{H - E_{1S} - \omega_1} \,
x^i \right| 2S \right> \,.
\end{eqnarray}
\end{subequations}
The perturbation (\ref{corrections}) 
leads to the following replacements, which include the 
first-order corrections to the various quantities
that are relevant to the $2S$ decay width,
\begin{eqnarray}
\label{quantities}
E_{1S} & \to & E_{1S} + \delta E_{1S} \,, \qquad
\delta E_{1S} = \left< 1S | \delta V | 1S \right>\,,
\nonumber\\[1ex]
E_{2S} & \to & E_{2S} + \delta E_{2S} \,, \qquad
\delta E_{2S} = \left< 2S | \delta V | 2S \right>\,,
\nonumber\\[1ex]
| 1S \rangle & \to & | 1S \rangle +
\left. \left. \left( \frac{1}{E_{1S} - H} \right)' \, \delta V 
\right| 1S \right>\,,
\nonumber\\[1ex]
| 2S \rangle & \to & | 2S \rangle +
\left. \left. \left( \frac{1}{E_{2S} - H} \right)' \, \delta V 
\right| 2S \right>\,,
\nonumber\\[1ex]
\omega_2 & \to & \omega_2 + \delta \omega_2 \,, \qquad
\delta \omega_2 = \delta E_{2S} - \delta E_{1S}\,.
\end{eqnarray}
The latter correction ensures that a perturbed energy 
conservation condition is fulfilled,
\begin{equation}
\omega_1 + \omega_2 + \delta \omega_2 = 
E_{2S} - E_{1S} + (\delta E_{2S} - \delta E_{1S})\,,
\end{equation}
i.e.~that the two photon frequencies add up to the 
perturbed transition frequency.

The first-order self-energy
correction $\delta \Gamma$ to the two-photon decay rate may 
be expressed as
\begin{equation}
\label{corrlength}
\frac{\delta \Gamma}{B} = 2\, \int\limits_0^{\omega_{\rm max}}
d\omega_1 \, \omega^3_1 \, \omega^3_2 \, \zeta \, \delta \zeta +
3 \, \delta \omega_2 \, \int\limits_0^{\omega_{\rm max}}
d\omega_1 \, \omega^3_1 \, \omega^2_2 \, \zeta^2 \,,
\end{equation}
where the correction $\delta \zeta$ is the sum of six terms,
\begin{equation}
\delta \zeta = \sum_{j=1}^6 \delta \zeta_j\,,
\end{equation}
to be defined as follows, and the second term 
on the right-hand side of (\ref{corrlength}) is due to 
perturbed energy conservation condition. 
The quantity $B$ may be inferred from 
(\ref{tpd1}), (\ref{standard}) and (\ref{VLamb}) as
\begin{equation}
\label{defB}
B = \frac{16}{81} \, \frac{\alpha^3}{\pi^2} \, \ln[(Z\alpha)^{-2}]\,.
\end{equation}
The terms $\delta \zeta_1$ and $\delta \zeta_2$ 
are related to energy perturbations to the matrix elements,
\begin{widetext}
\begin{subequations}
\label{defdeltazeta}
\begin{eqnarray}
\label{defdeltazeta1}
\delta \zeta_1 &=& \left< 1S \left| x^i \, 
\left( \frac{1}{H - E_{2S} + \omega_1} \right)^2 \,
x^i \right| 2S \right>\, 
\left< 2S \left| \delta V \right| 2S \right>\,,
\\[1ex]
\label{defdeltazeta2}
\delta \zeta_2 &=& \left< 1S \left| \delta V \right| 1S \right>\,
\left< 1S \left| x^i \, 
\left( \frac{1}{H - E_{1S} - \omega_1} \right)^2 \,
x^i \right| 2S \right>\,, 
\end{eqnarray}
whereas the terms $\delta \zeta_{3,4,5,6}$ are perturbations to the 
initial- and final-state wave functions, 
\begin{eqnarray}
\label{defdeltazeta3}
\delta \zeta_3 &=& \left< 1S \left| x^i \, 
\frac{1}{H - E_{2S} + \omega_1} \,
x^i \, \left( \frac{1}{E_{2S} - H} \right)' \delta V
\right| 2S \right>\,,
\\[1ex]
\label{defdeltazeta4}
\delta \zeta_4 &=& \left< 1S \left| x^i \, 
\frac{1}{H - E_{1S} - \omega_1} \,
x^i \, \left( \frac{1}{E_{2S} - H} \right)' \delta V
\right| 2S \right>\,,
\\[1ex]
\label{defdeltazeta5}
\delta \zeta_5 &=& \left< 1S \left| 
\delta V \, \left( \frac{1}{E_{1S} - H} \right)' \,
x^i \, \frac{1}{H - E_{2S} + \omega_1} \,
x^i \right| 2S \right>\,,
\\[1ex]
\label{defdeltazeta6}
\delta \zeta_6 &=& \left< 1S \left| 
\delta V \, \left( \frac{1}{E_{1S} - H} \right)' \,
x^i \, 
\frac{1}{H - E_{1S} - \omega_1} \,
x^i \right| 2S \right>\,.
\end{eqnarray}
\end{subequations}
\end{widetext}

%
%
\section{VELOCITY GAUGE}
\label{velocity}

We now discuss the evaluation of radiative corrections
in the velocity gauge, where the interaction Hamiltonian 
is given by
\begin{equation}
H'_{\rm int} = -e \, \frac{\bm{p} \cdot \bm{A}}{m} + 
e^2 \, \frac{\bm{A}^2}{2 m^2}.
\end{equation}
According to (\ref{tpd1}) and (\ref{tpd2}),
the leading-order decay rate in the velocity gauge is
\begin{equation}
\label{decayvel}
\frac{\Gamma'_0}{A} = \int\limits_0^{\omega_{\rm max}}
d\omega_1 \, \omega_1 \, \omega_2 \, \xi^2\,,
\end{equation}
where $A$ is defined in (\ref{defA}),
$\omega_2 \equiv E_{2S} - E_{1S} - \omega_1$ and 
$\omega_{\rm max} \equiv E_{2S} - E_{1S}$.
The quantity $\xi$ is the sum of two terms,
\begin{subequations}
\label{defxi}
\begin{equation}
\xi = \xi_1 + \xi_2\,,
\end{equation}
where
\begin{eqnarray}
\label{defxi1}
\xi_1 &=& \left< 1S \left| \frac{p^i}{m} \, 
\frac{1}{H - E_{2S} + \omega_1} \,
\frac{p^i}{m} \right| 2S \right> \,,
\nonumber\\[1ex]
\label{defxi2}
\xi_2 &=& \left< 1S \left| \frac{p^i}{m} \, 
\frac{1}{H - E_{1S} - \omega_1} \,
\frac{p^i}{m} \right| 2S \right> \,.
\end{eqnarray}
\end{subequations}
Gauge invariance of the leading-order decay-rate 
[see Eqs.~(\ref{decaylength}) and (\ref{decayvel})]
\begin{equation}
\label{leadingGamma}
\Gamma_0 = \Gamma'_0
\end{equation}
immediately follows from Eq.~(\ref{tpd2});
this equation may be rewritten in a compact form as
\begin{equation}
\label{easy}
\xi = - \omega_1 \, \omega_2 \, \zeta\,.
\end{equation}
Equation (\ref{easy}) may be proven easily by repeated application of 
the commutator relation(s)
\begin{equation}
\frac{p^i}{m} = 
{\rm i}\, [H - E_{2S} + \omega_1, x^i] =
{\rm i}\, [H - E_{1S} - \omega_1, x^i] \,.
\end{equation}

Now the first-order correction to the two-photon decay rate,
in the velocity gauge, is
\begin{equation}
\label{corrvel}
\frac{\delta \Gamma'}{B} = 2\, \int\limits_0^{\omega_{\rm max}}
d\omega_1 \, \omega_1 \, \omega_2 \, \xi \, \delta \xi +
\delta \omega_2 \, \int\limits_0^{\omega_{\rm max}}
d\omega_1 \, \omega_1 \, \xi^2 \,,
\end{equation}
where the prime denotes the velocity-gauge form of the 
correction and $B$ is defined in (\ref{defB}). 
We desire to show that $\delta\Gamma = \delta\Gamma'$.

The correction $\delta \xi$ finds a natural 
representation as the sum of eight terms,
\begin{equation}
\delta \xi = \sum_{j=1}^8 \delta \xi_j\,.
\end{equation}
In analogy to (\ref{defdeltazeta1}) and (\ref{defdeltazeta2}), 
$\delta \xi_1$ and $\delta \xi_2$ are energy perturbations,
\begin{widetext}
\begin{subequations}
\label{defdeltaxi}
\begin{eqnarray}
\label{defdeltaxi1}
\delta \xi_1 &=& \left< 1S \left| \frac{p^i}{m} \, 
\left( \frac{1}{H - E_{2S} + \omega_1} \right)^2 \,
\frac{p^i}{m} \right| 2S \right>\, 
\left< 2S \left| \delta V \right| 2S \right>\,,
\\[1ex]
\label{defdeltaxi2}
\delta \xi_2 &=& \left< 1S \left| \delta V \right| 1S \right>\,
\left< 1S \left| \frac{p^i}{m} \, 
\left( \frac{1}{H - E_{1S} - \omega_1} \right)^2 \,
\frac{p^i}{m} \right| 2S \right>\, .
\end{eqnarray}
The terms $\delta \xi_{3,4,5,6}$ are perturbations to the 
initial- and final-state wave functions, 
\begin{eqnarray}
\label{defdeltaxi3}
\delta \xi_3 &=& \left< 1S \left| \frac{p^i}{m} \, 
\frac{1}{H - E_{2S} + \omega_1} \,
\frac{p^i}{m} \, \left( \frac{1}{E_{2S} - H} \right)' \delta V
\right| 2S \right>\,,
\\[1ex]
\label{defdeltaxi4}
\delta \xi_4 &=& \left< 1S \left| \frac{p^i}{m} \, 
\frac{1}{H - E_{1S} - \omega_1} \,
\frac{p^i}{m} \, \left( \frac{1}{E_{2S} - H} \right)' \delta V
\right| 2S \right>\,,
\\[1ex]
\label{defdeltaxi5}
\delta \xi_5 &=& \left< 1S \left| 
\delta V \, \left( \frac{1}{E_{1S} - H} \right)' \,
\frac{p^i}{m} \, \frac{1}{H - E_{2S} + \omega_1} \,
\frac{p^i}{m} \right| 2S \right>\,,
\\[1ex]
\label{defdeltaxi6}
\delta \xi_6 &=& \left< 1S \left| 
\delta V \, \left( \frac{1}{E_{1S} - H} \right)' \,
\frac{p^i}{m} \, 
\frac{1}{H - E_{1S} - \omega_1} \,
\frac{p^i}{m} \right| 2S \right>\,.
\end{eqnarray}
Finally, $\delta \xi_{7,8}$ are due to the seagull term,
\begin{eqnarray}
\label{defdeltaxi7}
\delta \xi_7 &=& -\frac{3}{m} \,
\left< 1S \left| \left( \frac{1}{E_{2S} - H} \right)' \,
\delta V \right| 2S \right>\,,
\\[1ex]
\label{defdeltaxi8}
\delta \xi_8 &=& -\frac{3}{m} \,
\left< 1S \left| \delta V \, 
\left( \frac{1}{E_{1S} - H} \right)' \right| 2S \right>\,.
\end{eqnarray}
\end{subequations}
%

%
%
\section{PROOF OF GAUGE INVARIANCE}
\label{proof}

Here, we merely present the results of the analysis carried out in 
detail in App.~\ref{appa}.
Indeed, using Eqs.~(\ref{deltaxi1tolength})---(\ref{deltaxi6tolength}),
as well as (\ref{rel1}) and~(\ref{rel2}),
we obtain the compact relation
\begin{eqnarray}
\label{invariance}
\delta \xi &=& - \omega_1 \, \omega_2 \, \delta \zeta - 
\delta\omega_2 \,  \omega_1 \, \zeta \,.
\end{eqnarray}
In view of this relation,
we can rewrite (\ref{corrlength}) and (\ref{corrvel})
using (\ref{invariance}),
\begin{eqnarray}
\frac{\delta \Gamma'}{B} &=& 2\, \int\limits_0^{\omega_{\rm max}}
d\omega_1 \, \omega_1 \, \omega_2 \, \xi \, \delta \xi +
\delta \omega_2 \, \int\limits_0^{\omega_{\rm max}}
d\omega_1 \, \omega_1 \, \xi^2 
\nonumber\\[1ex]
&=& 2 \, \int\limits_0^{\omega_{\rm max}}
d\omega_1 \, \omega_1 \, \omega_2 \, (-\omega_1 \, \omega_2 \, \zeta) \, 
\left[ - \omega_1 \, \omega_2 \, \delta \zeta - 
\delta\omega_2 \, \omega_1 \, \zeta \right] 
+ \delta \omega_2 \, \int\limits_0^{\omega_{\rm max}}
d\omega_1 \, \omega^3_1 \, \omega^2_1 \, \zeta^2 
\nonumber\\[1ex]
&=& 2 \, \int\limits_0^{\omega_{\rm max}}
d\omega_1 \, \omega^3_1 \, \omega^3_2 \, \zeta \, \delta \zeta 
+ (2 + 1) \, \delta \omega_2 \, \int\limits_0^{\omega_{\rm max}}
d\omega_1 \, \omega^3_1 \, \omega^2_1 \, \zeta^2 
\nonumber\\[1ex]
&=& \frac{\delta \Gamma}{B} \,.
\end{eqnarray}
This proves the gauge invariance 
$\delta \Gamma = \delta \Gamma'$ of the 
logarithmic radiative corrections to the two-photon
decay rate of the metastable $2S$ state in hydrogenlike 
systems. The gauge invariance 
of the leading-order decay rate ($\Gamma_0 = \Gamma'_0$) has been 
indicated in Eq.~(\ref{leadingGamma}).
\end{widetext}

%
%
\section{NUMERICAL RESULTS}
\label{numerical}

({\em Leading order.})
We recall that, according to (\ref{tpd1}),
the well-known leading-order nonrelativistic effect
$\Gamma_0$ is of the order of $\alpha^2\,(Z\alpha)^6$.
The result for the two-photon decay 
width of the metastable $2S$ state is
\begin{equation}
\Gamma_0 = 0.001\,318\,222 \,\,\alpha^2\, (Z \alpha)^6 \, m\,.
\end{equation}
This translates into 
\begin{subequations}
\label{resnr}
\begin{eqnarray}
\Gamma_0 &=& 8.229\,351\,997 \, Z^6\, {\rm s}^{-1}\\[1ex]
&=& 1.309\,742\,049 \, Z^6\, {\rm Hz}\,.
\end{eqnarray}
\end{subequations}

({\em Radiative correction.})
In view of Eqs.~(\ref{standard}) and (\ref{VLamb}),
the leading logarithmic radiative correction $\delta\Gamma$
is of the order of
\begin{equation}
\delta\Gamma \sim 
\alpha^3\,(Z\alpha)^8\,\ln[(Z\alpha)^{-2}]\,m\,,
\end{equation}
i.e.~of relative order $\alpha\,(Z\alpha)^2\,\ln[(Z\alpha)^{-2}]$
with respect to $\Gamma_0$.
In the length gauge, 
the relevant expression for $\delta \Gamma$
can be found in Eq.~(\ref{corrlength}).
[For clarity, 
we would like to indicate that 
the correction $\delta \omega_2$ occurring in 
the expression (\ref{corrlength})
is defined in (\ref{quantities}), the quantity $\zeta$
can be found in (\ref{defzeta}),
and the terms $\delta\zeta_i$ ($i = 1,\dots,6$) are defined 
in Eq.~(\ref{defdeltazeta}).]
In the velocity gauge, 
the relevant expression for $\delta \Gamma'$ 
can be found in (\ref{corrvel}), with the 
$\delta\xi_i$ ($i = 1,\dots,8$) being defined 
in Eq.~(\ref{defdeltaxi}).

According to (\ref{corrlength}) and (\ref{corrvel}),
both $\delta \Gamma$ as well as $\delta \Gamma'$
find a natural representation as the sum of 
two terms, the first of which summarizes the 
perturbations to the matrix elements, and the 
second is a consequence of the perturbed energy conservation
condition for the transition.
Gauge invariance $\delta \Gamma = \delta \Gamma'$ has 
been shown in Sec.~\ref{proof}, yet it is 
instructive to observe that there are indeed
considerable cancellations among the 
two contributions to $\delta \Gamma$ and $\delta \Gamma'$.
Specifically, we have from the first and the second 
terms on the right-hand sides of (\ref{corrlength})
and (\ref{corrvel}), respectively,
\begin{eqnarray}
\frac{\delta \Gamma}{\Gamma_0} &=&
(29.542-31.562) \,
\frac{\alpha}{\pi}\,(Z\alpha)^2\, 
\ln[(Z\alpha)^{-2}] \,,\\[1ex]
\frac{\delta \Gamma'}{\Gamma'_0} &=&
(8.500-10.521)\,\frac{\alpha}{\pi}\,(Z\alpha)^2\, 
\ln[(Z\alpha)^{-2}]\,.
\end{eqnarray}
[The cancellations appear to be typical for 
radiative corrections to decay rates; this has recently
been observed in connection with radiative
corrections to the {\em one}-photon decay
of $P$ states~\cite{SaPaCh2004}.]

The final result for the leading logarithmic 
correction to the decay width of the metastable
$2S$ state is
\begin{equation}
\label{result}
\frac{\delta \Gamma}{\Gamma_0} =
\frac{\delta \Gamma'}{\Gamma_0} =
-2.020\,536\, \frac{\alpha}{\pi}\,(Z\alpha)^2\,\ln[(Z\alpha)^{-2}]\,.
\end{equation}
The calculation of $\delta \zeta$ and $\delta \xi$ 
involves expressions analogous to those encountered 
in~\cite{Je2003jpa}. In~\cite{KaIv1997c}, 
the coefficient has been given as $-2.025(1)$, 
which is in agreement with the current calculation.
[There is a misprint in the overall sign of the 
correction as given in the abstract of~\cite{KaIv1997c};
one should follow the sign indicated in Eq.~(8) {\em ibid.}]

The result (\ref{result}), converted to Hertz 
and/or inverse seconds, reads
\begin{subequations}
\label{resultHertz}
\begin{eqnarray}
\delta \Gamma &=& -3.273 \times 10^{-7}\, Z^8 \, 
\ln[137^2\,Z^{-2}] \, {\rm Hz} 
\\[1ex] 
&=& -2.057 \times 10^{-6}\, Z^8 \, 
\ln[137^2\,Z^{-2}] \, {\rm s}^{-1}\,.
\end{eqnarray}
\end{subequations}
For low $Z$,
the highly suppressed $M1$ one-photon decay 
$2S \to 1S$ is numerically smaller than the radiative
correction (\ref{resultHertz}) to the two-photon decay
(see Refs.~\cite{BrTe1940,BeSa1957,Dr1971,FeSu1971,Jo1972})
because it lacks the large logarithm:
\begin{equation}
\Gamma_{M1} = 2.496 \times 10^{-6} Z^{10} \, {\rm s}^{-1}\,.
\end{equation}
All results indicated in this article for $\Gamma$ and
$\delta \Gamma$ relate to the metastable $2S$ state; however
the approach may easily be generalized to the two-photon
decay of other states.

%
%
\section{CONCLUSIONS}
\label{conclusions}

In the current investigation,
the derivation of the leading radiative correction to the
two-photon decay width of the metastable $2S$ state
in hydrogenlike atoms has been based on the
effective ``radiative potential'' (\ref{radpot})
discussed in Sec.~\ref{radcor}.
It has been shown that the gauge invariance 
of the corrections holds due to the interplay of 
corrections to the transition matrix elements
on the one hand and corrections due to perturbed
energy conservation conditions on the other hand
[first and second terms on the right-hand sides
of~(\ref{corrlength}) and (\ref{corrvel}),
respectively]. The corrections to the transition matrix elements
are again divided into corrections to the wave function
(these were referred to as the {\em d}-terms in~\cite{KaIv1997c}),
and to the energies that enter into the propagator denominators,
which were termed {\em f} in the length-gauge
calculation~\cite{KaIv1997c}.
The length- and velocity-gauge forms
of the correction are discussed in Secs.~\ref{length}
and~\ref{velocity}. 
The gauge invariance of the radiative correction holds
(even) on the level of the effective treatment as implied by the 
radiative potential (\ref{radpot}),
as shown in Sec.~\ref{proof}.
All derivations are presented in some detail, for clarity and transparency.
The numerical evaluation in Sec.~\ref{numerical} follows
immediately.

There are two more results of the current paper, probably
of rather minor importance, which should only
briefly be mentioned: first of all, the relativistic
result (without radiative corrections)
for the decay rate at $Z=1$ has previously been indicated
as $8.229\,{\rm s}^{-1}$~\cite{GoDr1981,PaJo1982,Dr1986,DeJo1997},
whereas in~\cite{ShBr1959}, the
(nonrelativistic) result has been
indicated as $8.226 \pm 0.001$ inverse seconds. The current investigation
[Eq.~(\ref{resnr})] confirms that the discrepancy has been due to
a certain overestimation of
the numerical accuracy in the early nonrelativistic
calculation~\cite{ShBr1959}, not due
to a conceivable large relativistic shift.
Second, the discussion in Sec.~\ref{leading} clarifies that the concept of
a decay width as an imaginary part of a self-energy~\cite{BaSu1978}
generalizes to the two-loop self-energy shift, in which case
the imaginary part gives rise
to the two-photon decay width.
 
The leading-order nonrelativistic
contribution to the 
two-photon decay width is of the 
order of $\alpha^2 (Z\alpha)^6\,m\,c^2$ (see Sec.~\ref{leading}).
The self-energy radiative correction to the two-photon decay is
of the order of $\alpha^3 \, (Z\alpha)^8 \, \ln[(Z\alpha)^{-2}]\,m\,c^2$,
as discussed in Sec.~\ref{numerical},
with explicit results indicated in Eqs.~(\ref{result})
and~(\ref{resultHertz}). It would be interesting to evaluate 
also the constant term of relative order $\alpha \, (Z\alpha)^2$.
This term supplements the 
logarithm evaluated here which is of relative order
$\alpha \, (Z\alpha)^2 \, \ln[(Z\alpha)^{-2}]$.
According to our experience, in bound-state calculations,
the nonlogarithmic, constant term has an 
opposite sign as compared to the leading logarithm, 
and its magnitude is two or three times larger than the 
coefficient of the logarithm.
This is true for radiative 
corrections~\cite{SaPaCh2004} as well as Lamb-shift 
effects~\cite{SaYe1990,remarkZ}.

One should note a rather general interest in various 
intriguing details related to the two-photon 
decay process, which are not restricted to 
the search for conceivable parity admixtures to the $2S$ state
(see e.g.~\cite{Mo1978,DuEtAl1997}).
Although accurate measurements of integrated decay rates
are difficult~\cite{DuEtAl1999}, there is some hope that 
in low-$Z$ and middle-$Z$ ionic systems, 
experiments will eventually profit from 
the possibilities offered by electron-beam ion traps,
especially when combined with conceivable x-ray lasers
that could be used in order to excite the trapped ions
into the metastable states.

Finally, we recall
that accurate measurements of the two-photon decay
width test the $2S$ state for parity-violating $2P$-admixtures and
can therefore be used as a test for a conceivable electron
or nuclear (electric) dipole moment or for interactions
via ``anapole'' or ``pseudocharge'' currents~\cite{Ze1957,Ze1959,SaFe1966}.
One particularly interesting investigation on
hydrogenlike Ar$^{17+}$, with an elucidating discussion
of the issues related to parity admixtures,
has been given in~\cite{MaSc1972}.

\acknowledgments

The author acknowledges elucidating discussions with 
Holger Gies and Wilhelm Becker on questions related to 
the gauge invariance, and insightful conversations
with Krzysztof Pachucki
regarding quantum electrodynamic effects in bound systems.
The author wishes to thank 
Gordon Drake for very helpful remarks.
Sabine Jentschura is acknowledged for carefully reading the
manuscript.
The stimulating atmosphere at the National Institute
of Standards and Technology has contributed to 
the completion of this project.

\appendix

\begin{widetext}

%
%
\section{RELATIONS AMONG MATRIX ELEMENTS}
\label{appa}

In this appendix, we present in detail the relations needed
for the proof of the identity (\ref{invariance}).
For $\delta\xi_1$ as defined in (\ref{defdeltaxi1}), we have
\begin{subequations}
\begin{eqnarray}
\label{deltaxi1tolength}
& & \left< 1S \left| \frac{p^i}{m} \, 
\left( \frac{1}{H - E_{2S} + \omega_1} \right)^2 \,
\frac{p^i}{m} \right| 2S \right> =
- \omega_1 \, \omega_2 \,
\left< 1S \left| x^i \, \left( \frac{1}{H - E_{2S} + \omega_1} \right)^2 \,
x^i \right| 2S \right> 
\nonumber\\[1ex]
& & \quad + (\omega_2 - \omega_1) \, 
\left< 1S \left| x^i \, \frac{1}{H - E_{1S} + \omega_1} \,
x^i \right| 2S \right> +
\left< 1S \left| x^i \, x^i \right| 2S \right>\,.
\end{eqnarray}
We notice the term $\delta \zeta_1$ emerge on the 
right-hand side [see Eq.~(\ref{defdeltazeta1})].
The corresponding relation for $\delta\xi_2$ reads
\begin{eqnarray}
\label{deltaxi2tolength}
& & \left< 1S \left| \frac{p^i}{m} \, 
\left( \frac{1}{H - E_{1S} - \omega_1} \right)^2 \,
\frac{p^i}{m} \right| 2S \right> =
- \omega_1 \, \omega_2 \,
\left< 1S \left| x^i \, \left( \frac{1}{H - E_{1S} - \omega_1} \right)^2 \,
x^i \right| 2S \right>
\nonumber\\[1ex]
& & \quad + (\omega_1 - \omega_2) \, 
\left< 1S \left| x^i \, \frac{1}{H - E_{1S} - \omega_1} \,
x^i \right| 2S \right> +
\left< 1S \left| x^i \, x^i \right| 2S \right>\,.
\end{eqnarray}
For $\delta \xi_3$, the following relation is useful,
\begin{eqnarray}
\label{deltaxi3tolength}
\lefteqn{ \left< 1S \left| \frac{p^i}{m} \, 
\frac{1}{H - E_{2S} + \omega_1} \,
\frac{p^i}{m} \, \left( \frac{1}{E_{2S} - H} \right)' \delta V
\right| 2S \right> =
- \omega_1 \, \omega_2 \,
\left< 1S \left| x^i \, \frac{1}{H - E_{2S} + \omega_1} \,
x^i \left( \frac{1}{E_{2S} - H} \right)' \delta V \right| 2S \right> }
\nonumber\\[1ex]
& & - \omega_2 \, 
\left< 1S \left| x^i \, \frac{1}{H - E_{2S} + \omega_1} \,
x^i \right| 2S \right> \, \left< 2S \left| \delta V \right| 2S \right>
+ \left< 1S \left| x^i \, (H - E_{2S} + \omega_2) \,
x^i \left( \frac{1}{E_{2S} - H} \right)' \delta V \right| 2S \right> 
\nonumber\\[1ex]
& & - \left< 1S \left| x^i \, x^i \right| 2S \right>\,
\left< 2S \left| \delta V \right| 2S \right>\,.
\end{eqnarray}
For $\delta \xi_4$, we have
\begin{eqnarray}
\label{deltaxi4tolength}
\lefteqn{ \left< 1S \left| \frac{p^i}{m} \, 
\frac{1}{H - E_{1S} - \omega_1} \,
\frac{p^i}{m} \, \left( \frac{1}{E_{2S} - H} \right)' \delta V
\right| 2S \right> =
- \omega_1 \, \omega_2 \,
\left< 1S \left| x^i \, \frac{1}{H - E_{1S} - \omega_1} \,
x^i \left( \frac{1}{E_{2S} - H} \right)' \delta V \right| 2S \right> }
\nonumber\\[1ex]
& & - \omega_1 \, 
\left< 1S \left| x^i \, \frac{1}{H - E_{1S} - \omega_1} \,
x^i \right| 2S \right> \, \left< 2S \left| \delta V \right| 2S \right>
+ \left< 1S \left| x^i \, (H - E_{1S} - \omega_2) \,
x^i \left( \frac{1}{E_{2S} - H} \right)' \delta V \right| 2S \right> 
\nonumber\\[1ex]
& & - \left< 1S \left| x^i \, x^i \right| 2S \right>\,
\left< 2S \left| \delta V \right| 2S \right>\,.
\end{eqnarray}
The term $\delta \xi_5$ may be reformulated according to
\begin{eqnarray}
\label{deltaxi5tolength}
\lefteqn{ \left< 1S \left| \delta V \, \left( \frac{1}{E_{1S} - H} \right)'\,
\frac{p^i}{m} \, \frac{1}{H - E_{2S} + \omega_1} \,
\frac{p^i}{m} \, \right| 2S \right> =
- \omega_1 \, \omega_2 \,
\left< 1S \left| \delta V \, \left( \frac{1}{E_{1S} - H} \right)'\,
x^i \, \frac{1}{H - E_{2S} + \omega_1} \, x^i \right| 2S \right> }
\nonumber\\[1ex]
& & + \omega_1 \, \left< 1S \left| \delta V \right| 1S \right> \,
\left< 1S \left| x^i \, \frac{1}{H - E_{1S} - \omega_1} \,
x^i \right| 2S \right> 
+ \left< 1S \left| \delta V \, \left( \frac{1}{E_{1S} - H} \right)'\,
x^i \, (H - E_{1S} - \omega_1) \, x^i \right| 2S \right> 
\nonumber\\[1ex]
& & - \left< 1S \left| \delta V \right| 1S \right>\,
\left< 1S \left| x^i \, x^i \right| 2S \right>\,.
\end{eqnarray}
Finally, we have for $\delta \xi_6$
\begin{eqnarray}
\label{deltaxi6tolength}
\lefteqn{ \left< 1S \left| \delta V \, \left( \frac{1}{E_{1S} - H} \right)'\,
\frac{p^i}{m} \, \frac{1}{H - E_{1S} - \omega_1} \,
\frac{p^i}{m} \, \right| 2S \right> =
- \omega_1 \, \omega_2 \,
\left< 1S \left| \delta V \, \left( \frac{1}{E_{1S} - H} \right)'\,
x^i \, \frac{1}{H - E_{1S} - \omega_1} \, x^i \right| 2S \right> }
\nonumber\\[1ex]
& & + \omega_2 \, \left< 1S \left| \delta V \right| 1S \right> \,
\left< 1S \left| x^i \, \frac{1}{H - E_{1S} - \omega_1} \,
x^i \right| 2S \right> 
+ \left< 1S \left| \delta V \, \left( \frac{1}{E_{1S} - H} \right)'\,
x^i \, (H - E_{2S} + \omega_1) \, x^i \right| 2S \right> 
\nonumber\\[1ex]
& & - \left< 1S \left| \delta V \right| 1S \right>\,
\left< 1S \left| x^i \, x^i \right| 2S \right>\,.
\end{eqnarray}
\end{subequations}
However, the relations (\ref{deltaxi1tolength})---(\ref{deltaxi6tolength})
are not yet sufficient in order to proceed with the proof of 
gauge invariance. We also need
\begin{subequations}
\begin{eqnarray}
\label{rel1}
\lefteqn{\left< 1S \left| x^i \, [(H - E_{1S}) + (H - E_{2S})] \, x^i \, 
\left( \frac{1}{E_{2S} - H} \right)'\, \delta V \right| 2S \right> }
\nonumber\\[1ex]
&=& \frac{3}{m} \, 
\left< 1S \left| \left( \frac{1}{E_{2S} - H} \right)' \,
\delta V \, \right| 2S \right> 
+ \left< 1S \left| x^i \, x^i \right| 2S \right> \,
\left< 2S \left| \delta V \right| 2S \right> \,,
\\[1ex]
\label{rel2}
\lefteqn{\left< 1S \left| \delta V \, \left( \frac{1}{E_{1S} - H} \right)'\,
x^i \, [(H - E_{1S}) + (H - E_{2S})] \, x^i \, \right| 2S \right> }
\nonumber\\[1ex]
&=& \frac{3}{m} \,
\left< 1S \left| \delta V \,
\left( \frac{1}{E_{1S} - H} \right)' \right| 2S \right> 
+ \left< 1S \left| \delta V \right| 1S \right> \,
\left< 1S \left| x^i \, x^i \right| 2S \right> \,.
\end{eqnarray}
\end{subequations}
We notice the (negative of) the seagull terms (\ref{defdeltaxi7}) and 
(\ref{defdeltaxi8}) emerge.

\end{widetext}

\end{document}